\documentclass[pra,preprint,showpacs,groupedaddress,amsmath,amssymb]{revtex4}
\pdfoutput=1
\usepackage{latexsym}
\usepackage{amsfonts}
\usepackage{amsmath}
\usepackage{graphicx}
\usepackage{dcolumn}
\usepackage{bm}

\begin{document}

\title{Clusters of Exceptional Points for a Laser Control of Selective Vibrational Transfer}

\author{R. Lefebvre$^{1,2}$, A. Jaouadi$^{1,3}$ and O. Atabek$^{1*}$}
\affiliation{$^1$Institut des Sciences Mol\'eculaires d'Orsay, B\^at. 350, Universit\'e Paris-Sud 11 (CNRS),  91405 Orsay France\\
$^2$U.F.R. de Physique Fondamentale et Appliqu\'ee, Universit\'e Pierre et Marie Curie, 75321 Paris, France\\
$^3$Laboratoire Aim\'e Cotton, CNRS, B\^at. 505, Universit\'e Paris-Sud 11, 91405 Orsay, France}
\email[O. Atabek: ]{osman.atabek@u-psud.fr}

\date{\today}

\begin{abstract}
When a molecule is exposed to a laser field, all field-free vibrational states become resonances, with complex quasienergies calculated using Floquet theory. There are many ways to produce the coalescences of pairs of such quasienergies, with appropriate wavelength-intensity choices which define Exceptional Points (EP) in the laser parameter plane. We dress for the molecular ion H$_2^+$ an exhaustive map of these exceptional points which appear in clusters. Such clusters can be used to define several vibrational transfer scenarios implying more than a single exceptional point, exchanging single or multiple vibrational quanta. The ultimate goal is molecular vibrational cooling by transfering an initial (thermal, for instance) population on a final (ground, for instance) single  vibrational state. 
\end{abstract}


\maketitle

\section{Introduction}

Molecular cooling at ultracold temperatures down to the quantum degenerate regime opens the possibility to study small clusters in Bose-Einstein condensates, by the preselection of quantum states and their deterministic manipulation, exerting control on their internal couplings. There are several techniques to create and manipulate well-defined ro-vibrational states of diatomic molecules involved in these clusters. Photo-association is one of them. For a system like Na$_2$, a good candidate for the formation of translationnaly cold molecules, very high vibrational levels close to the dissociation limit are prepared, from which selective transfers may lead, through some laser purification schemes, to the obtention of a single ro-vibrational level (ultimately the ground one). Gerardo Delgado-Barrio's most outstanding contributions that are honoured through the present issue of \emph{Chemical Physics}, precisely concern small molecular clusters and their long-range van der Waals interactions leading to an accurate determination of the positions and lifetimes of highly excited vibrational states, as the ones obtained by photo-association (see for instance \cite{Gerardo} and references therein).

In the way of performing vibrational cooling in diatomic species like Na$_2$, Cs$_2$ or Rb$_2$, prepared by photo-association \cite{Jones} and presenting a high density of states, we are discussing in this paper, one of the major ingredients of the purification scheme: the laser control of the vibrational state-to-state transfers based on the coalescence of pairs of resonance energies described using the Floquet formalism. The illustration is given on a lighter diatomic, namely  H$_2^+$. Resonances are complex eigenenergies of the system, their imaginary parts being related to their fragmentation rate. If it is possible, by a particular choice of the laser parameters controlling a system, to produce a degeneracy of two such energies, the corresponding point in the (wavelength, intensity) parameter plane  is called an exceptional point (EP) \cite{Kato1,Heiss1}.  A very important property of exceptional points is that, by an adiabatic variation of the parameters along a closed contour around an EP, it is  possible to go from one resonance state  to another \cite{hernandez,Heiss2}. We have recently proposed such a laser control scenario for a transfer from one field-free vibrational state to another, based on a single EP, in the case of molecular photo-dissociation \cite{PRL}. The first step is to find a specific loop in the laser parameter plane such that, starting from a given field-free vibrational state $v$,  a single resonance  adjusts continuously its characteristics (energy and width) to reach a different vibrational state $v \pm 1$.  In a second step we introduce the time it takes to follow the loop to fulfill the conditions for an adiabatic transfer. This finally leads to an estimate of the percentage of undissociated molecules at the end of the pulse. For H$_2^+$, we showed that about 10 percent of the molecules could survive in the most favorable conditions, which turns out to be the adiabatic following of exclusively Feshbach resonances (with rather long lifetimes) as will be discussed later. This control strategy corresponds to a complete vibrational transfer between neighbouring field-free vibrational states ($\Delta v=\pm1$). By complete we mean that the initial population of a given vibrational state $v$ is selectively transferred to a single final field-free vibrational state (either $v+1$ or $v-1$) with no population left on $v$. A transfer involving distant neighbours (with $\arrowvert \Delta v \arrowvert > 1$) can be considered by referring to an ordered sequence of EPs, bringing the system from $v+n$ to $v$, by steps of single vibrational quantum changes ($\Delta v=-1$) as has previously been suggested by us \cite{PRL}. The purpose of the present paper is to show the existence of multiple EPs in the parameter plane and to define laser pulse shapes taking advantage of these EPs for a thorough control of vibrational transfers going beyond $\Delta v=\pm1$. The ultimate goal is molecular cooling using a control scheme to obtain a single ro-vibrational state, by successive transfers starting, for instance, from a thermal distribution of initial vibrational states.

The paper is organized as follows: Section II gives the elements of the Floquet formalism used in this study. The method to obtain approximate values of the coordinates of EPs from a coincidence condition is described in Section III, together with an accurate calculation of the complete set of EPs. Section IV considers several types of adiabatic pulses based on the use of more than a single EP, differing in that from most studies presented so far. Each scenario is checked through the Floquet adiabatic theory in order to obtain the amount of undissociated molecules at the end of the pulse. 

\section{The Floquet formalism}

 We are referring to a rotationless one-dimensional model involving only two electronic states $1s \sigma_g$ (noted as $\arrowvert g \rangle$) and $2p \sigma_u$ (noted as $\arrowvert u \rangle$) whose potential energy curves are displayed in Fig. 1 after a photon dressing mechanism explained hereafter. 
 The wave function of the system as expanded on these two states is written 
\begin{equation}
\label{eq:1}
 \arrowvert \Psi (R,t) \rangle = \chi_g(R,t)\arrowvert g \rangle +
 \chi_u (R,t) \arrowvert u \rangle\,,
\end{equation}
\noindent where $R$ is the nuclear coordinate, $\chi_g$, $\chi_u$ are unknown functions describing field-assisted nuclear dynamics.
The frozen rotation assumption is validated by considering the long rotational periods of H$_2^+$ (estimated as tens of ps) with respect to the pulse durations under consideration (less than 0.1 ps).

The molecule interacting with a linearly polarized laser of leading frequency $\omega$, is described by a periodic time-dependent Hamiltonian (ideally valid for a cw field) and
the solution of the time-dependent Schr\"odinger equation (TDSE) can be written as 
\begin{eqnarray}
\left[ \begin{array}{c} \chi_g (R,t) \\
\chi_u (R,t) \end{array}\right] = e^{-iE_Ft/\hbar} \left[ \begin{array}{c} \phi_g (R,t) \\
\phi_u (R,t) \end{array}\right]\,.
\end{eqnarray}
\noindent 
The periodicity in time of $\phi_k(R,t)$ ($k=g,u$) allows us to expand these functions in Fourier series
\begin{equation}
\phi_k(R,t)=\sum_{n=-\infty}^{+\infty} e^{in\omega t}\varphi_{k}^{n}(R)\,.
\end{equation}
Finally, adopting the length-gauge and the long-wavelength approximation for the matter-field coupling, the Fourier components $\varphi_{k}^{n}(R)$ are given as solutions of the coupled differential equations \cite{Book}
\begin{equation}
\Big[ T +
V_{g,u}(R)+n\hbar\omega-E_F\Big]\varphi_{g,u}^{n}(R)-1/2~ \mathcal{E}_0~ \mu(R)\Big[
\varphi_{u,g}^{n-1}(R) + \varphi_{u,g}^{n+1}(R)\Big]=0
\end {equation}
\noindent $T=(-\hbar^2/(2\,\mathcal{M}))(d^2\hspace*{-1.00mm}/dR^2)$ is the usual
vibrational kinetic energy operator and $\mathcal{M}$ is the reduced mass. $V_g(R)$ and $V_u(R)$ are the two Born-Oppenheimer potentials.  $\mu(R)$ is the electronic transition dipole moment between states $\arrowvert g \rangle$ and $\arrowvert u \rangle$.
 $\mathcal{E}_0$ is the laser electric field peak amplitude, with a wavelength $\lambda= 2 \pi c /\omega$ and intensity $I\propto \mathcal{E}_0^{2}$.
\noindent Solutions with Siegert outgoing wave boundary conditions in the open channels \cite{Book} produce  complex quasienergies of the form
 $E_F=E_R-i\Gamma_R/2$, where $\Gamma_R$ is the decay rate.
 The equations are solved with a matching technique based on the Fox-Goodwin propagator \cite{FOX}, with exterior complex scaling \cite{moiseyev}. The single photon processes are described with only two channels, that is $\varphi_g^{0}(R)$ and $\varphi_u^{-1}(R)$. Figure 1 displays for H$_2^+$ the various field-dressed potentials involved in the Floquet picture. The left panel shows the so-called diabatic potentials, the dotted potentials being those involved in the single photon absorption process, while the others below correspond to two- and three-photon absorptions. The ones above describe virtual emission of photons. The right panel shows the two adiabatic potentials V$_+(R)$ and V$_-(R)$ obtained from the diagonalization, at each value of $R$, of the potential matrix describing single photon absorption. 

\begin{figure}[hbt]
\begin{center}
\includegraphics[width=12.5cm]{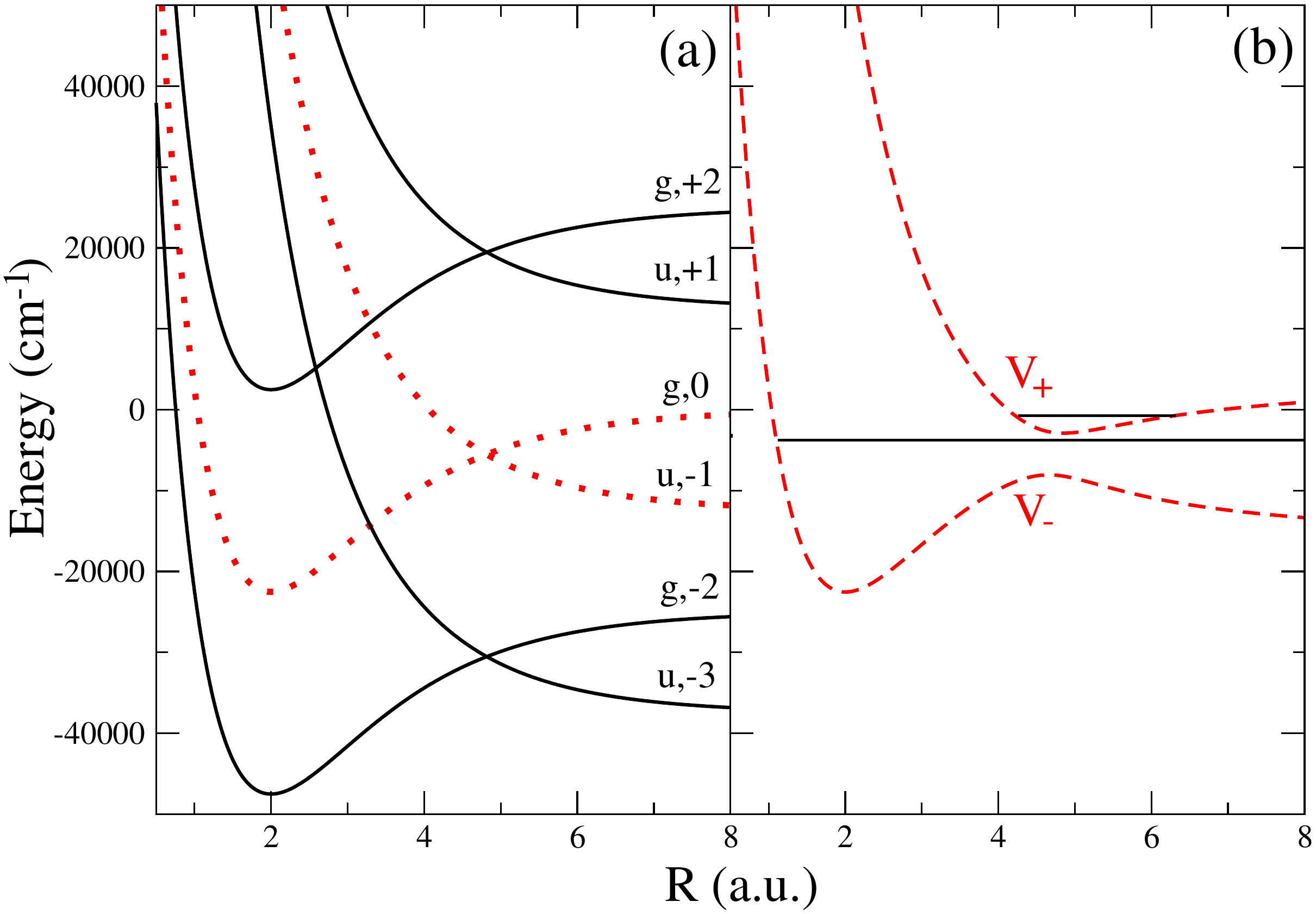}
\caption{(color online)(a): The potentials of the two electronic states ($g$ and $u$) of H$_2^+$ dressed by the photon quanta $\hbar \omega = 0.056$ u.a. The dotted potentials describe  single photon absorption. (b):The adiabatic potentials produced by the diagonalization of the potential matrix with a field intensity of $10^{13} W/cm^2$. The horizontal lines are representatives of a shape resonance of the lower adiabatic potential and of a bound state of the upper adiabatic potential.}
\end{center}
\end{figure}

\section{Localization of exceptional points} 

Adiabatic potentials are very useful guides for the interpretation of laser-induced resonances.
We have previously observed \cite{JPC} that the Floquet resonances fall into two categories, those which are of Feshbach type (F), and those which are essentially of shape nature (S). The resonances belonging to the first class (F) are, at the lowest order of radiative coupling approximation, interpreted as the bound states of the V$_+(R)$ potential interacting with the continuum of the $V_-(R)$. Increasing the intensity results into an increasing of their energies, as the adiabatic potentials split further and further. As for their widths, a decreasing behavior is obtained, due to lesser non-adiabatic couplings with the continuum of the V$_-(R)$ potential. Opposite to this behavior, the energies of shape resonances (S) associated, at zeroth order, with the resonances of the V$_-(R)$ potential decrease, whereas their widths increase with intensity, due to potential barrier lowering. An important consequence is that the energy of a bound state of the upper adiabatic potential is always below that of the associated Floquet energy. This is understandable when considering the coupling of such a bound state with the continuum of states of the lower adiabatic potential. The energy of such a state is necessarily above the region of the avoided crossing. The coupling of this state with the continuum of the lower adiabatic potential is therefore showing a maxinum for an energy close to the energy of the avoided crossing since this is the region of maximum dependance of the electronic and nuclear wave functions versus the nuclear coordinate.  The consequence is that the resonance level shift to go from the bound state to the Floquet state is positive. This property can be turned into a method to get approximate values for the parameters of EPs. We show in Figure 2  how the adiabatic bound energies change with the wavelength when the intensity is very small (\emph{i.e.}, about $10^3$W/cm$^2$), just enough to produce an avoided crossing between the two potentials. Also shown as horizontal lines are the diabatic energies at such an intensity: these are simply the vibrational energies of the field-free molecule. The variation range for the wavelength ($\lambda$ from about 110 nm up to 900 nm) is such that we are only dealing with diabatic curve-crossings occuring for internuclear distances larger than the equilibrium position of the attractive potential (the so-called c$_+$ case in the language of spectroscopists when dealing with predissociation \cite{Bandrauk}). The diagnostic to estimate roughly the wavelength at which an EP is expected is as follows. 

\begin{figure}[hbt]
\begin{center}
\includegraphics[width=12.5cm]{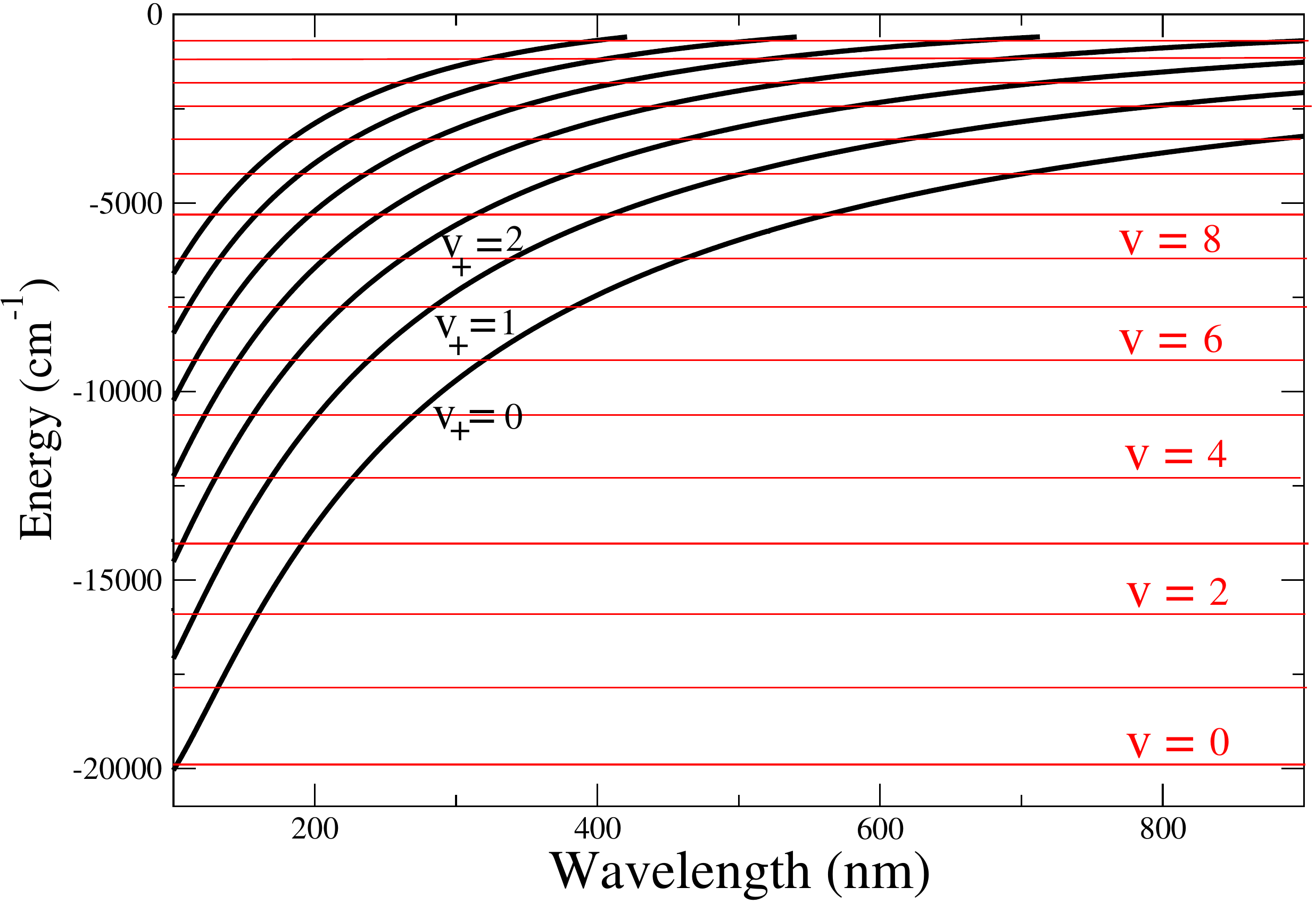}
\caption{(color online) The diabatic energies (as horizontal red thin lines for $v=0$ up to $v=15$) and the energies of the quasi-bound states of the upper adiabatic potential at an intensity just sufficient to provoke an avoided crossing (black solid curves for $v_+=0$ up to $v_+=6$) as a function of wavelength. Every crossing between curves of the two families gives an approximate value for the wavelength of an exceptional point: such a crossing is associated with a change of the character of the resonance state: Feshbach on the left of the crossing, shape on the right.}
\end{center}
\end{figure}

\noindent Consider for example the crossing between the diabatic level $v=12$ with the adiabatic level with quantum number $v_+=1$ occuring for $\lambda \sim 789.7$ nm. If the wavelength is smaller than 
789.7 nm, the adiabatic level is below the diabatic energy (which is in fact the Floquet energy for such a field intensity). An intensity increase will have as a consequence a positive energy shift for the adiabatic level $v_+=1$ which will progressively merge with the Floquet energy. On the other hand if $\lambda$ is larger than  789.7 nm, this merging is no longer possible. In other words in the first case we are facing a resonance of Feshbach type, while in the other case the resonance is of shape type. The next step is to start from these values of the wavelength and to examine the energies and widths as a function of field intensity. Based on the approximate findings of Figure 2, we illustrate in Figure 3, accurate solutions of Eq.(4) for $v=12$ and $v=13$. Some adjustment is needed until the situation of interchange between Feshbach and shape characters is produced. Finally, the transition from Feshbach to shape is well illustrated showing the behaviour of both energies and widths in the immediate neighbourhood of an EP. Two slightly different values of the wavelength (below or above the EP, respectively) produce an interchange in the classification. In the left panel $v=13$ is shape (increase of width, decrease of energy with increasing intensity) while $v=12$ is Feshbach (decrease of width, increase of energy). In the right panel, it is just the reverse: $v=12$ is shape, while $v=13$ is Feshbach.  A typical morphological signature of an EP emerges for a wavelength slightly below (within 4-digits accuracy): a crossing for the energies and a tweezer-like profile for the widths, as a function of intensity.

\begin{figure}[hbt]
\begin{center}
\includegraphics[width=12.5cm]{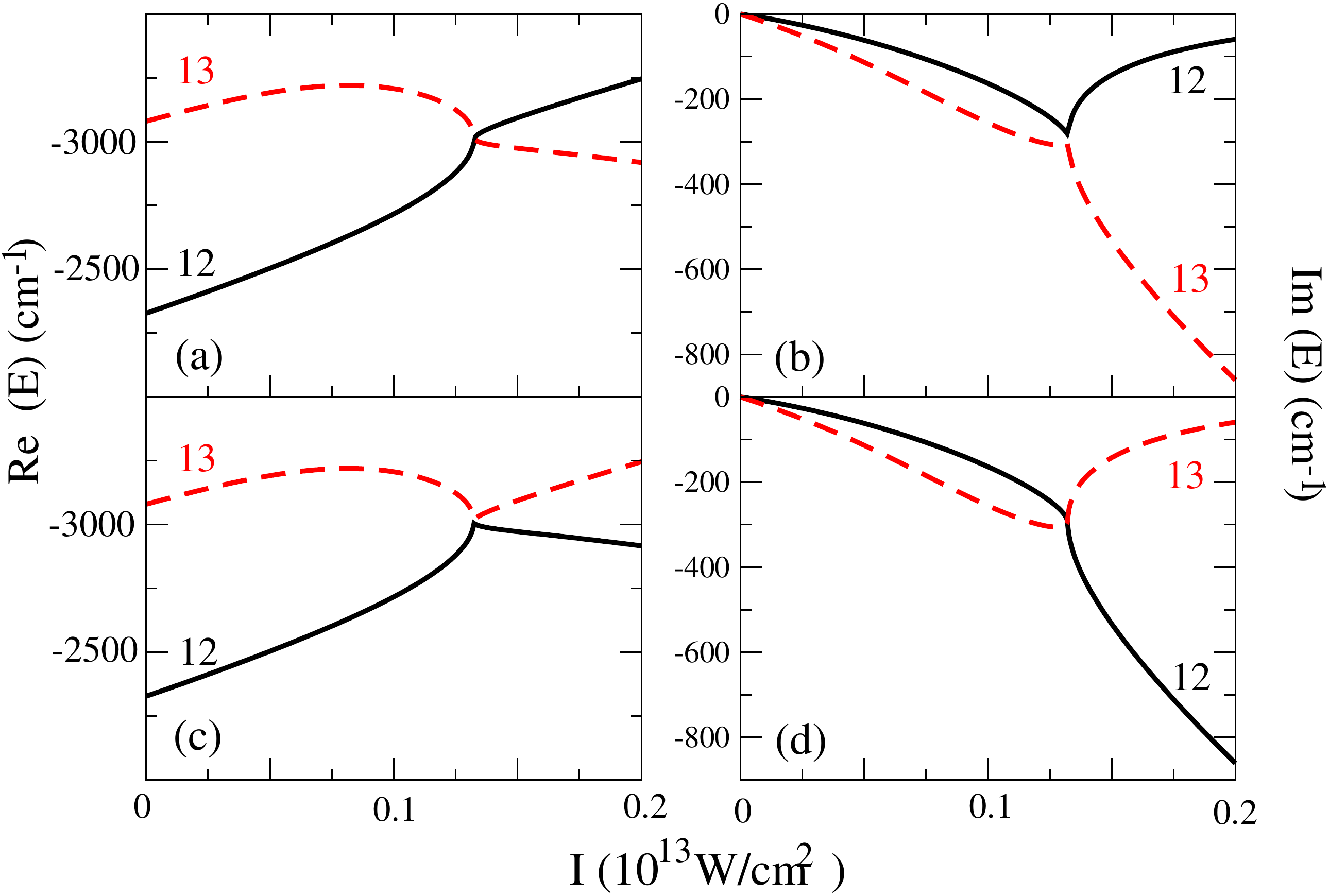}
\caption{(color online) Real and imaginary parts of Floquet resonances $v=12, 13$ as a function of the laser intensity illustrating the interchange of the shape versus Feshbach characters of a resonance when very close to an EP. First row corresponds to a wavelength of 788.2 nm, whereas the second is for 788.3 nm (slightly below the approximate value of 789.7 nm predicted in Figure 2). In the upper panels $v=12$ is Feshbach, while $v=13$ is shape. In the lower panels there is an inversion. A Feshbach resonance has an energy which increases for increasing intensities and a decreasing width. For a shape resonance the energy decreases, while the width increases.}
\end{center}
\end{figure}

The results displayed in Figure 3 constitute a full check of the presence of an EP for a pair of vibrational levels (taken as an example to be $v=12$ and $13$) as roughly predicted by the network of wavelength-dependent diabatic and adiabatic energy levels of Figure 2. It is, in particular, worthy noting that, for a given pair of resonances, one may expect several EPs occuring at different wavelengths, as suggested by the several possible crossings between the given diabatic energies and successive adiabatic levels $v_+=0,1,2,...$ An important issue for the control scenarios to be discussed in the following Section is to refer to a complete map of the EPs. Such a map for H$_2^+$ is displayed in Figure 4. Two interesting features are depicted: (i) Higher wavelengths ($\lambda > 400 ~$ nm) give rise to curve-crossings in higher energy regions corresponding to a higher density of levels, as is clear from Figure 2. The consequence is that, field strengths with lower intensities (I $< 0.5 ~ 10^{13}$ W/cm$^2$) are enough to produce, through moderate  energy shifts, the resonance coalescence required for an EP to occur. A typical example is displayed in the inset (a) of Figure 4, involving high energy pairs $v=12$ to $16$. Lower wavelengths ($\lambda < 400 ~$ nm) on the contrary, induce curve-crossings in lower energy regions, with much less density of levels, necessitating higher intensities (I $> 0.5~ 10^{13}$ W/cm$^2$) for the coalescence condition. Such a typical case, for low energy pairs $v=7$ to $10$, is illustrated in the inset (b) of Figure 4. (ii) The EPs are organized in clusters (as the ones emphazised in the insets) which are well ordered in terms of the initial field-free vibrational levels giving rise to them; EPs originating from higher field-free vibrational states being obtained for lower wavelengths and stronger fields. 

\begin{figure}[hbt]
\begin{center}
\includegraphics[width=12.5cm]{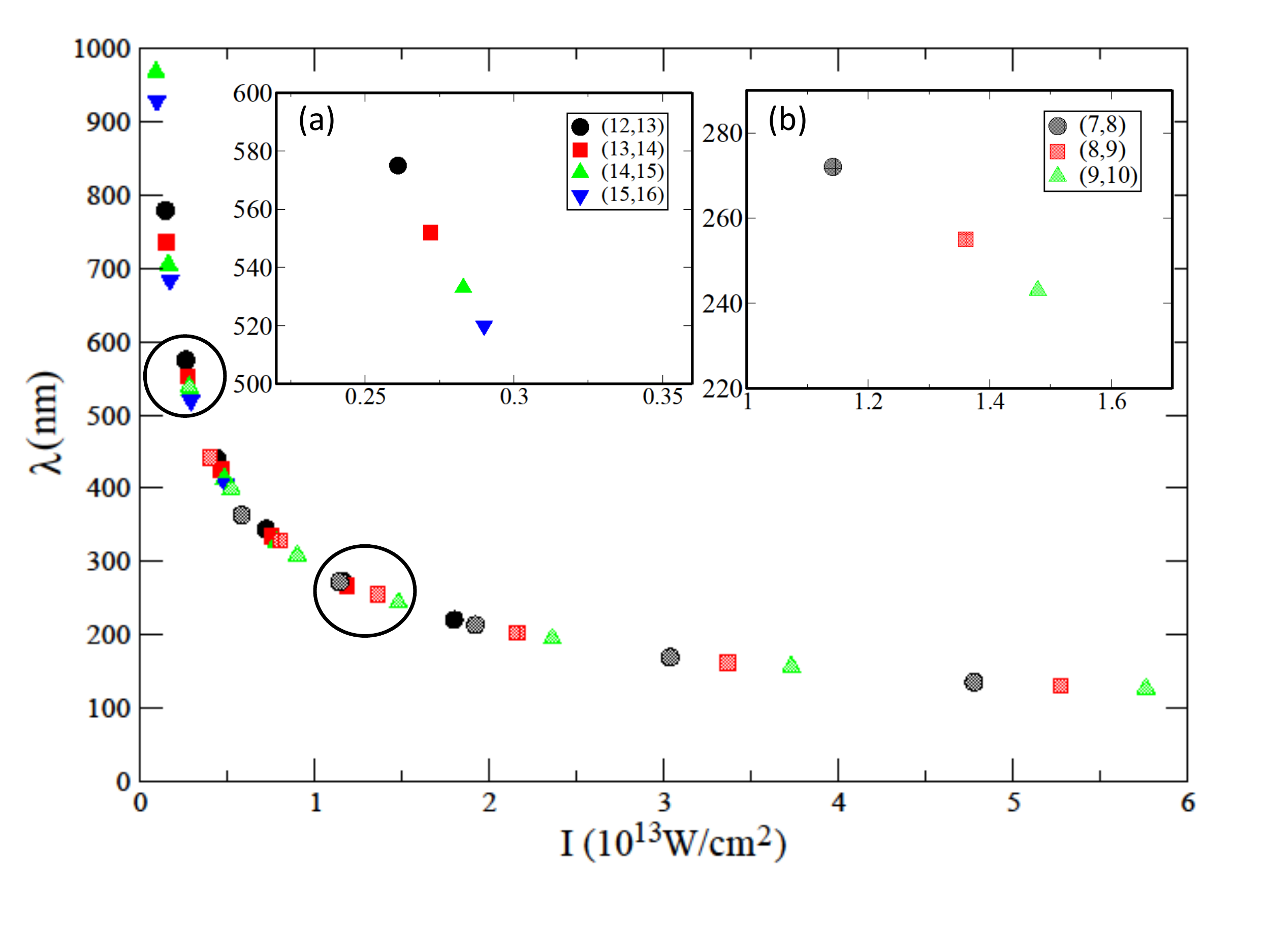}
\caption{(color online)The set of exceptional points of H$_2^+$ derived from the levels up to $v=16$ and a minimum wavelength producing a crossing to the right of the equlibrium point of the ground state electronic potential. The two insets represent clusters of EPs corresponding to different wavelength regimes and concerning different pairs of levels (as indicated in parenthesis). Note that the ordering of all clusters follows the same rule: EPs originating from higher field-free vibrational states being localized at lower wavelengths and stronger fields.}
\end{center}
\end{figure}

\begin{figure}[hbt]
\begin{center}
\includegraphics[width=12.5cm]{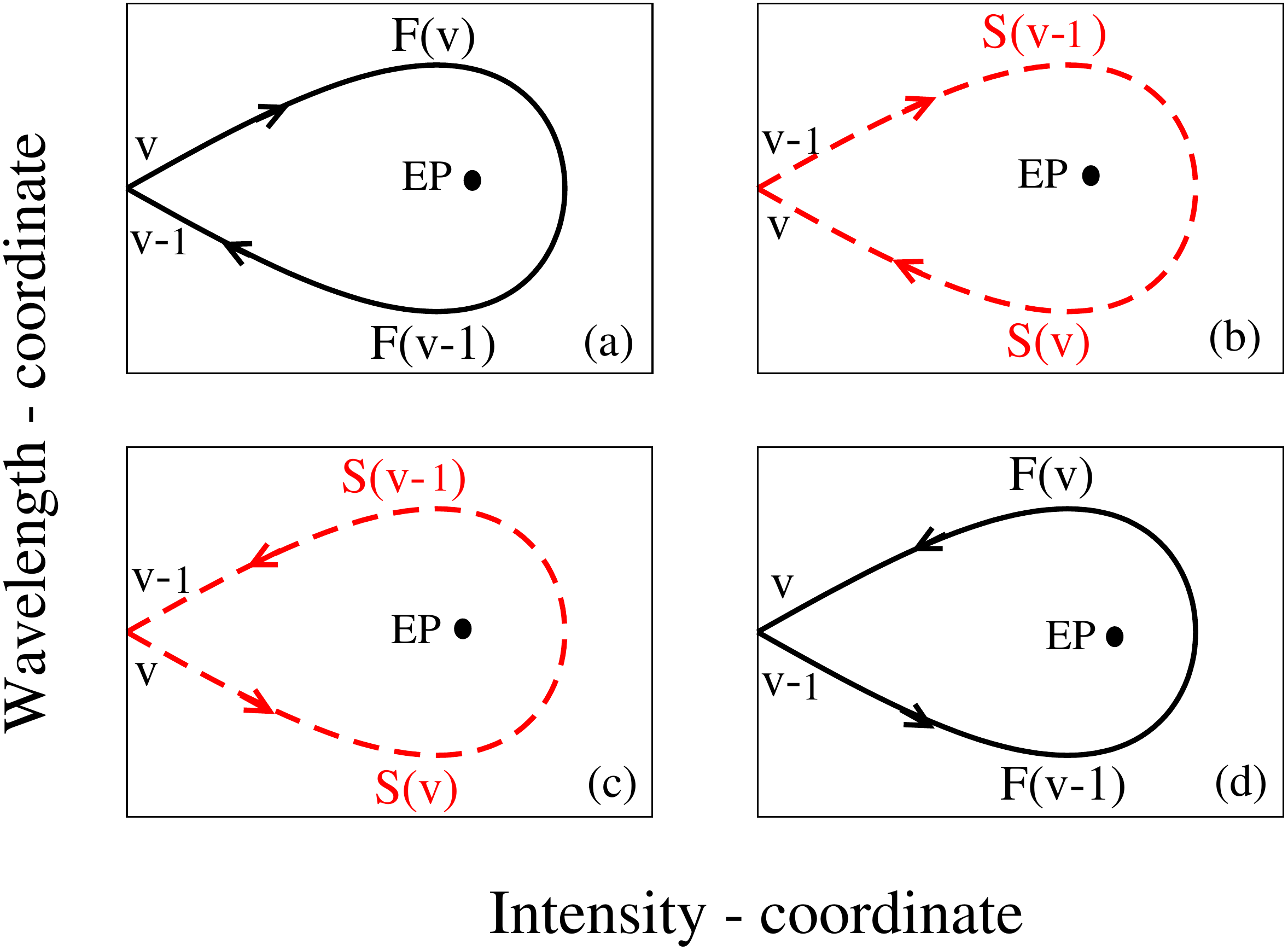}
\caption{(color online)The various ways to change the vibrational quantum by one unit. Two of them ((a) and (d)) follow all along the loop a Feshbach resonance, while with (b) and (c) it is a shape resonance which is followed.}
\end{center}
\end{figure}

\section{Transfers based on clusters of exceptional points}

It is well known by now \cite{Heiss1,hernandez} that a laser pulse shaped such as to produce a contour in the parameter plane (wavelength and intensity) which encircles an EP, leads to an interchange of the resonance energies involved in the coalescence process, and thus to a vibrational transfer. The control strategy consists in applying this chirped pulse which \emph{selectively} transfers the probability density from one field-free vibrational level to another. As for the \emph{efficiency} of the transfer, it critically depends on the shape or Feshbach nature of the resonances. Based on this classification, we schematically show in Figure 5 the various loops which can encircle an EP and produce a transfer from state $v$ to $v-1$, or $v-1$ to $v$, with $\Delta v = \pm 1$. An important observation is that a loop around the EP cannot affect the nature of a resonance state: a shape resonance remains shape (panels (b) and (c)), while a Feshbach one remains Feshbach (panels (a) and (d)). 
As is clear from Figure 5(a,c), there are two possibilities to end up in state $v-1$ starting from $v$, (\emph{i.e.}; $\Delta v = -1$). The clock-wise contour (panel (a)) follows the Feshbach resonance $F(v)$ originating from $v$ and switches, after encircling the corresponding EP($v, v-1$) onto the Feshbach resonance $F(v-1)$ which ultimately merges in the vibrational level $v-1$ when the pulse is off. The reverse is observed for the anticlock-wise contour (panel(c)) where the shape resonance $S(v)$ matches $S(v-1)$. If the transfer which is looked for consists in reaching $v$, starting from $v-1$  (\emph{i.e.}, $\Delta v = 1$) there is again the possibility to follow Feshbach type resonances all along the loop, but taking opposite contour directions: Anticlock-wise for Feshbach (panel (d)) or clock-wise for shape (panel(b)).
The purpose of the operation being to keep as high as possible the fraction of undissociated molecules, it is more efficient to follow a Feshbach resonance with a lower rate of dissociation rather than a shape one with its high rate.

Up to now, we have considered vibrational transfers involving but single quantum exchanges ($\Delta v =  \pm 1$), exploiting the branching possibilities offered by a single EP. In the following we wish to take advantage of the occurence, for different laser wavelengths and intensities, of multiple EPs as shown in Figure 4, to control cascade type transfers from $v+n$ to $v$ or from $v$ to $v-n$ (\emph{i.e.}; $\Delta v = \pm n$) using a single chirped pulse encircling several EPs between successive resonances.
Our example is based on the cluster of four EPs implying levels $v=12$ to $v=16$ shown as the insert (a) of Figure 4, with corresponding wavelength/intensity parameters collected in Table I. The loops which are in consideration for the transfers are of the form

\begin{equation}
I=I_{max}~\sin(\phi/2),~~~~~~\lambda=\lambda_{0}+\delta \lambda~ \sin(\phi)\,,
\end{equation}

\noindent with the angle $\phi$ taking a set of values between $0$ and $2 \pi$.
 
\begin{figure}[hbt]
\begin{center}
\includegraphics[width=12.5cm]{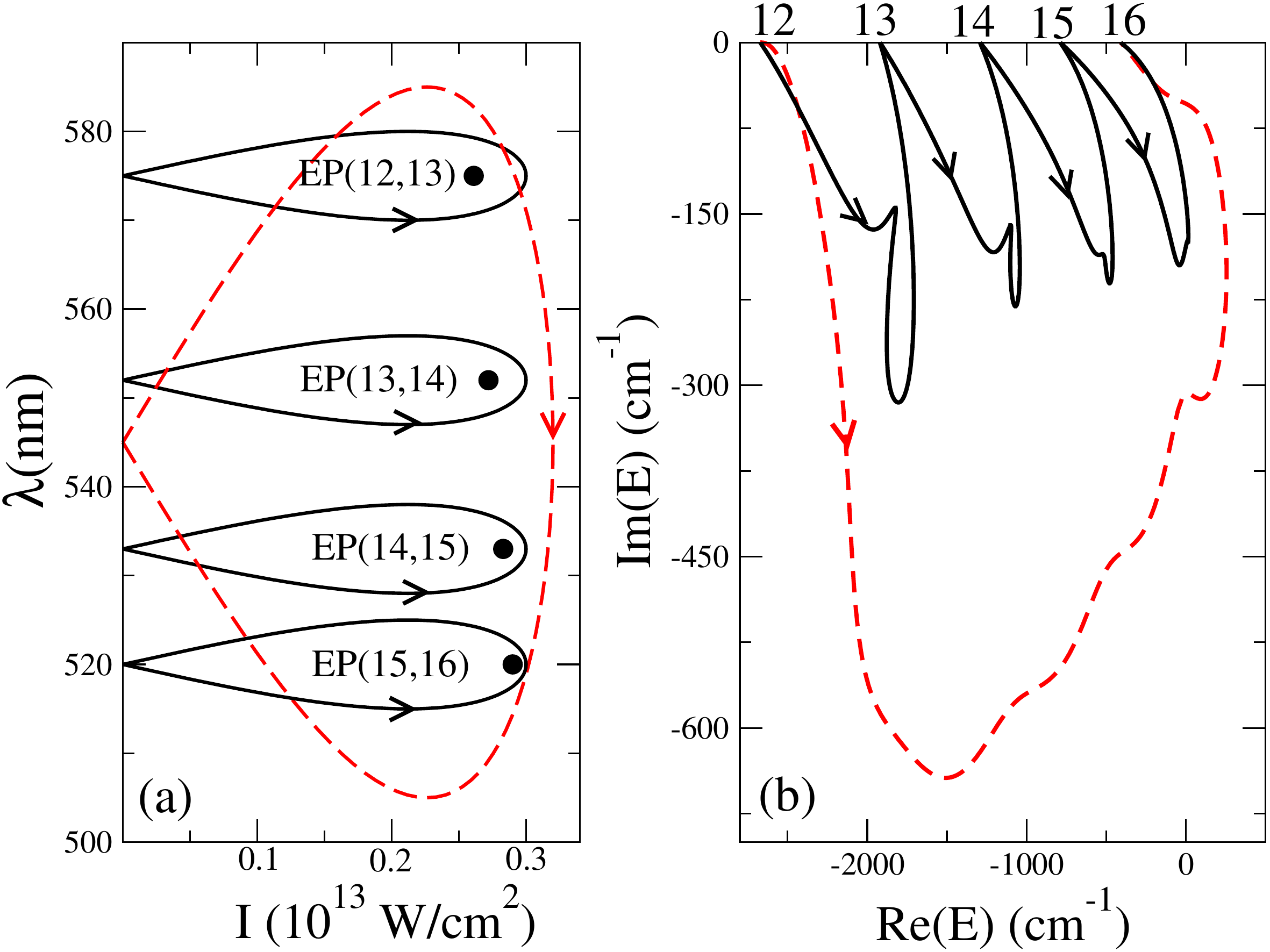}

\caption{(color online) (a) Laser loops in the (wavelength/intensity) parameter plane encircling EP($v, v+1$) for $v=12$ to $15$. The solid black contours correspond to the successive transfers strategy, the red dotted contour is for the cascade-type transfers strategy. (b) Corresponding resonance transfers in the complex energy plane (imaginary versus real parts of the complex energies).}
\end{center}
\end{figure}

Figure 6a illustrates the two strategies which are considered: (i) A series of successive laser pulses shaped as to encircle a single EP($v, v+1$) at a time; (ii) A single pulse encircling the whole cluster of the four EPs. Table I collects the parameters of the four successive loops chosen in order to encircle EP($v, v+1$) for $v=12$ to $15$, together with those of the single pulse encircling the cluster of the four EPs. 
With respect to the first strategy, starting from $v=12$ there are two ways to reach $v=13$, as is indicated in Figure 5: either by clock-wise (panel b) or anticlock-wise (panel d) contours. A similar observation applies to the other EPs involved in the cluster. Obviously, the most efficient scenario is the one which takes advantage of the presence of Feshbach-type resonances along the trajectories (\emph{i.e.}, anticlock-wise contours of Figure 5d) since it corresponds to the best trapping of the dissociation process. Actually, the dissociation rates of the Feshbach resonances involved in these transfers are less than 600 cm$^{-1}$, as compared with their shape analogues obtained using the clock-wise contour, with widths of the order of 2000 cm$^{-1}$. The resulting trajectories of the resonances (imaginary versus real parts of their complex energies) are displayed in Figure 6b.

The second strategy is based on cascade-type transfers, first from $v=12$ to $13$, followed by $v=13$ to $14$, and so on. Due to the specific molecular-dependent ordering of the EPs in the cluster (EP(12,13) being positionned at the highest wavelength) such transfers are only possible through the clock-wise contour of Figure 5b, involving thus shape-type resonances. 
 To understand this feature, let us follow in detail the transition from $v=12$ to $v=14$. The coordinates of the EP for the pair $v=12,13$ are $\lambda=575$ nm and $I=0.261$  (in units of $10^{13}$ W/cm$^2$) and for the pair $v=13,14$ $\lambda=552$ nm and $I=0.272$ (cf Table I). The loop defined on the second line of Table I has a wavelength reaching $584$ nm, higher than that of the first EP,  such that the associated resonance is of shape type. Once the transition to the resonance $v=13$ has been achieved for  $\lambda <575 $ nm the resonance approaching the second EP is still shape. It matches with $v=14$ again in the situation of panel (b) of Figure 5. The same reasoning applies to the following EPs involved in the cluster. The resulting trajectory is displayed in Figure 6b as a dotted line and exhibits dissociation rates (\emph{i.e.}, twice the imaginary parts of the energies) that may reach values as large as 1300 cm$^{-1}$.

To quantitatively compare the relative merits of the two strategies, with respect to their efficiency, we have to estimate the fraction of non-dissociated molecules left after the pulse is over. We make use of the adiabatic Floquet theory \cite{fleischer,ZWR}. This assumes that the chirped laser pulse envelope and its frequency vary sufficiently slowly with time such that the overall fraction of non-dissociated molecules $P_{ND}(t)$  at time $t$ is given by
\begin{equation} \label{P-undiss}
   P_{ND}(t) \; = \; \exp\left[-\hbar^{-1}~\int_0^{t}  \Gamma_R(t^{'})~dt^{'}\right] \;\; .
\end{equation}
Here, $\Gamma_R(t^{'})$ is associated with the relevant Floquet quasi-energy eigenvalue calculated using the instantaneous
field parameters at time $t^{'}$. The important information is  $P_{ND}(t_f)$ where $t_f$ is the laser pulse duration which has to be determined to satisfy an adiabatic transport condition. Adiabaticity refers to a transport of a field-free vibrational level on a single Floquet resonance, following it in time until it changes its label and end up into another vibrational level when the field is over. This requires a laser bandwidth narrow enough to avoid resonance mixing between the two-dimensional sub-spaces built on the pair of resonances participating to the EP and all the others \cite{Kato, TTND}. For the energy region of H$_2^+$ we are considering, $t_f > 30$fs roughly fulfills such a requirement. The results are displayed in Figure 7, for the two strategies. When considering the first scenario, the 30 fs laser pulse aiming in the ($\Delta v = 1$) transfer from $v=12$ to $13$, leaves a fraction of non-dissociated molecules calculated from Eq.(6) to be about $17~\%$. The second pulse, allowing the transfer from $v=13$ to $14$, leads to $1.4~\%$ of non-dissociated molecules. Finally, after four pulses, the system ends into the vibrational state $v=16$ ($\Delta v=4$), but with only $0.2~\%$ of stable molecules. This has to be compared with the fraction of non-dissociated molecules for the direct $v=12$ to $16$ transition, amounting to about $5~\%$ in the case of the single 30 fs laser pulse, encircling the cluster of EPs. The conclusion is that an upward ($\Delta v = 4$) multiple EP  transfer process, although based on shape resonances, is much more efficient as compared to the strategy of successively applied pulses even if these are following Feshbach resonances. A remaining interesting point is to whether a downward process ($\Delta v = -4$) be based on the use of a sequence of Feshbach resonances. As a result of the specific ordering of the EPs, at least for the case of H$_2^+$, this would involve the anticlock-wise contour displayed in Figure 5c, which once again, follows only shape-type resonances.

\begin{figure}[hbt]
\begin{center}
\includegraphics[width=12.5cm]{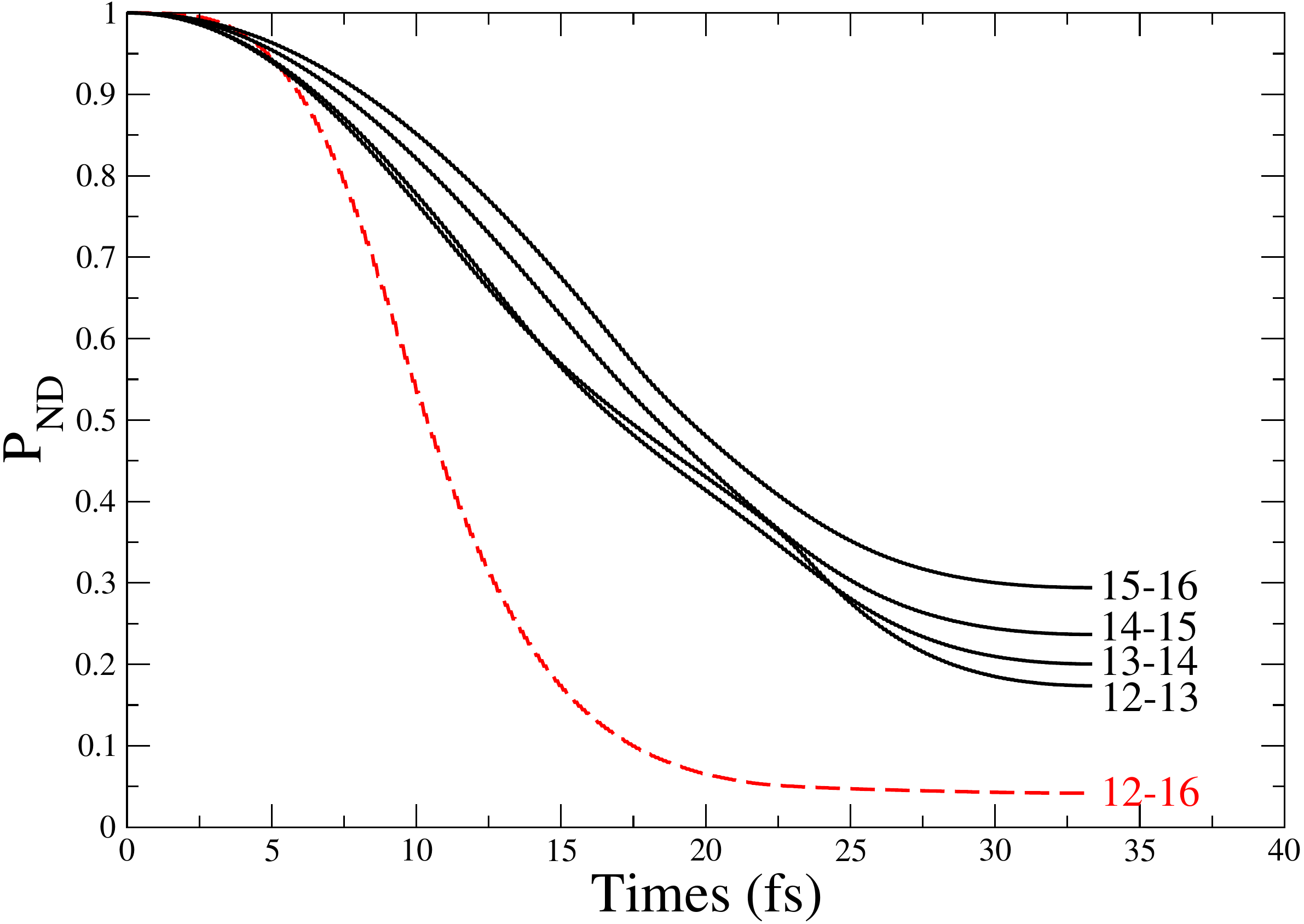}
\caption{(color online) Time evolution of the probability for the molecule to remain bound. The solid black lines represent single vibrational quantum transfers ($v$ to $v+1$). The dotted red line is for the direct 4-quanta transfer ($v=12$ to $16$).}
\end{center}
\end{figure}

Finally we are facing the question of the possible improvement of the second strategy (cascade-type transfers using clusters of EPs) as regard to its efficiency, that is its robustness with respect to dissociation. This could basically be done by exploiting clusters of EPs unfolded by weaker laser pulses. Resonances are then interchanging their labels at lower field intensities, for which the molecular system is better protected against dissociation. For H$_2^+$ a way to do this is to consider clusters of EPs located in the 700-800 nm (or even 900-1000 nm) wavelength range (see Figure 4). Laser pulses with more than two times weaker peak intensity may then be enough to encircle the EP cluster, offering much better robustness, through lower dissociation rates taking part in the exponentially decaying term of Eq.(6). For heavier molecular systems, like Na$_2$, we have recently shown that, due to much higher density of levels, resonance coalescences are obtained for much lower laser intensities. As a result the resonances participating in the adiabatic transfer process have widths more than ten times lower, leading to about 20 times better protection against dissociation \cite{PRLp}. Apart from this, and even more interestingly, the ordering of EPs for Na$_2$ in the wavelength/intensity plane turns out to be such that multiple EPs used in cascade-type transfers involve only Feshbach resonances, providing thus well improved robustness.

\begin{table}
\begin{center}
\begin{tabular}{p{2cm} p{2cm} p{2cm}| p{2cm} p{2cm} p{2cm}}
\hline
\hline
\multicolumn{3}{c|}{EP coordinates}& \multicolumn{3}{c}{Loop parameters}\\
\hline
v&     $\lambda$&    I &    $\lambda_0$&    $\delta  \lambda$& I$_{max}$ \\
\hline
12-13 & $575$&  $0.261$ & $575$&  $5$&  $0.30$ \\
13-14 & $552$&  $0.272$ & $552$&  $5$&  $0.30$ \\
14-15 & $533$&  $0.283$ & $533$&  $5$&  $0.30$ \\
15-16 & $520$&  $0.290$ & $520$&  $5$&  $0.30$ \\ \hline
12-16 &      &         &  $545$&  $40$&  $0.32$\\ \hline
\hline
\end{tabular}
\caption{The parameters of the five loops used to make the
calculations presented in Figure (6). The parameters are defined in Eq.(5).
Wavelengths are in nm and intensities in units of $10^{13}$ W/cm$^2$).}
\end{center}
\end{table}

\section{Conclusion}
We have shown that selective vibrational transfers are possible using several exceptional points in the laser-induced photo-dissociation of a molecular system with a rather simple experimental technique. This is an illustration that the existence of clusters of EPs in the parameter plane leads to a more complex structure in terms of multi-sheeted Riemann surfaces than that exhaustively studied when just one EP is present. While a single EP allows a transfer involving a single vibrational quantum, multiple EPs may lead to transfers with several exchanged quanta. The relative merits of two strategies are compared: either several pulses encicling a single EP at a time, monitoring thus the vibrational transfer between two adjacent resonances ($\Delta v = \pm 1$), or a single pulse to encircle several EPs between successive resonances allowing cascade transfers ($\Delta v = \pm n$) . This may be useful for the definition of purification schemes, with the goal to cool the molecule by obtaining a single ro-vibrational state. The proposed scheme can be applied to other molecular species such as Na$_2$, Cs$_2$ or Rb$_2$ which are good candidates for the formation of samples of translationnaly cold molecules \cite{Fatemi}. According to the specific molecular system in consideration, it is worth noting that EPs located at lower field intensities, or ordered in clusters offering multiple cascade transfers by following Feshbach (rather than shape) type resonances  can lead to much efficient population transfers with better protection against dissociation. Also more efficient laser control is expected by pulse shaping such as to take full advantage of the possible proximity of Zero Width Resonances \cite{ZWR} along the loop; a situation which seems to occur with heavier molecules and in particular in Na$_2$. Our group is actively pursuing investigations along these lines \cite{PRLp}.

\section{Acknowledgments}
OA gratefully acknowledges partial support from France-Canada CFQCU project 2010-19.


\begin{thebibliography}{100}
\bibitem{Gerardo} A. Vald\'es, R. Prosmiti, P. Villareal, G. Delgado-Barrio, D. Lemoine and B. Lepetit, J. Chem. Phys {\bf 126\rm}, 244314 (2007).
\bibitem{Jones} K. M. Jones, E. Tiesinga, P. D. Lett and P. S. Julienne, Rev. Mod. Phys. {\bf 78\rm}, 483 (2006).
\bibitem{Kato1} T. Kato, \textit{Perturbation Theory of Linear Operators} (Springer, Berlin, 1966).
\bibitem{Heiss1} W. D. Heiss,  Eur. Phys. J. D. {\bf 17\rm}, 1 (1999).
\bibitem{hernandez} E. Hern\'andez, A. J\'auregui and A. Mondrag\'on, J. Phys. A:Math. Gen. {\bf 39\rm}, 10087 (2006).
\bibitem{Heiss2} W. D. Heiss, Czech. J. Phys. {\bf 54\rm}, 54, 1091 (2004).
\bibitem{PRL} R. Lefebvre, O. Atabek, M. \u{S}indelka and N. Moiseyev, Phys. Rev. Lett.  {\bf 103\rm}, 123003 (2009).
\bibitem{Book}  O. Atabek, R. Lefebvre and T. T. Nguyen-Dang, in {\it Handbook of Numerical Analysis}, Vol. X, C. Le Bris Ed.; Elsevier: New York, 2003).
\bibitem{FOX} L. Fox and E. T. Goodwin, Phil. Trans. Roy. Soc. {\bf 245\rm}, 501(1953).
\bibitem{moiseyev} N. Moiseyev, Phys. Rep. {\bf 302\rm}, 212 (1998).
\bibitem{JPC} O. Atabek and R. Lefebvre, J. Phys. Chem. {\bf 114\rm}, 3031(2010).
\bibitem{Bandrauk} A. D. Bandrauk and J. P. Laplante, J. Chem. Phys. {\bf 65\rm}, 2592 (1976).
\bibitem{fleischer} A. Fleischer and N. Moiseyev, Phys. Rev. A {\bf 72\rm}, 032103 (2005).
\bibitem{ZWR} O. Atabek, R. Lefebvre, C. Lefebvre and T. T. Nguyen-Dang, Phys. Rev. A {\bf 77\rm}, 043413 (2008).
\bibitem{Kato} T. Kato, J. Phys. Soc. Jpn. {\bf 5\rm}, 435 (1950).
\bibitem{TTND} T. T. Nguyen-Dang, E. Sinelnikov, A. Keller and O. Atabek, Phys. Rev. A {\bf 76\rm}, 052118 (2007).
\bibitem{Fatemi} F. K. Fatemi, K. M. Jones, P. D. Lett, and E. Tiesinga, Phys. Rev. A {\bf 66\rm}, 053401(2002).
\bibitem{PRLp} O. Atabek, R. Lefebvre, M. Lepers, A. Jouadi, O. Dulieu, and V. Kokoouline, (submitted) arXiv:1011.2562.
\end{thebibliography}
\end{document}